\documentclass[twocolumn,aps,prl,floatfix,superscriptaddress]{revtex4-1}
\usepackage{amsmath,amssymb}
\usepackage{graphicx,float}
\usepackage{times,txfonts}
\usepackage{color}
\usepackage{multirow}
\usepackage{bbold}
\usepackage{color}
\usepackage{soul}
\usepackage{hyperref}
\usepackage{times,txfonts}
\usepackage{natbib}
\usepackage{nicefrac}
\usepackage{blindtext}
\usepackage[export]{adjustbox}

\newcommand{\abs}[1]{\left| #1 \right|}
\newcommand{\bra}[1]{\langle #1|}
\newcommand{\ket}[1]{| #1\rangle}
\newcommand{\braket}[2]{\left\langle {#1{\left| \vphantom{#1 #2} \right.} #2} \right\rangle}
\newcommand{\qo}[1]{``#1''}

\renewcommand{\epsilon}{\varepsilon}
\renewcommand{\phi}{\varphi}

\usepackage{sistyle}
\usepackage{soul}
\definecolor{lightblue}{RGB}{185,210,248}
\definecolor{YellowOrange}{RGB}{226,154,2}
\sethlcolor{YellowOrange}
\sethlcolor{lightblue}

\begin{document}
\title{Non-destructive measurement of an electron beam's orbital angular momentum}
\author{Hugo Larocque}
\affiliation{The Max Planck Centre for Extreme and Quantum Photonics, Department of Physics, University of Ottawa, 25 Templeton St., Ottawa, Ontario, K1N 6N5 Canada}
\author{Fr\'ed\'eric Bouchard}
\affiliation{The Max Planck Centre for Extreme and Quantum Photonics, Department of Physics, University of Ottawa, 25 Templeton St., Ottawa, Ontario, K1N 6N5 Canada}
\author{Vincenzo Grillo}
\affiliation{CNR-Istituto Nanoscienze, Centro S3, Via G Campi 213/a, I-41125 Modena, Italy}
\author{Alicia Sit}
\affiliation{The Max Planck Centre for Extreme and Quantum Photonics, Department of Physics, University of Ottawa, 25 Templeton St., Ottawa, Ontario, K1N 6N5 Canada}
\author{Stefano Frabboni}
\affiliation{CNR-Istituto Nanoscienze, Centro S3, Via G Campi 213/a, I-41125 Modena, Italy}
\affiliation{Dipartimento FIM Universit\'a di Modenae Reggio Emilia, Via G Campi 213/a, I-41125 Modena, Italy}
\author{Rafal E. Dunin-Borkowski}
\affiliation{Ernst Ruska-Centre for Microscopy and Spectroscopy with Electrons, Forschungszentrum J\"ulich, J\"ulich 52425, Germany}
\author{Miles J. Padgett}
\affiliation{School of Physics and Astronomy, Glasgow University, Glasgow, G12 8QQ, Scotland, UK}
\author{Robert W. Boyd}
\affiliation{The Max Planck Centre for Extreme and Quantum Photonics, Department of Physics, University of Ottawa, 25 Templeton St., Ottawa, Ontario, K1N 6N5 Canada}
\affiliation{Institute of Optics, University of Rochester, Rochester, New York, 14627, USA}
\author{Ebrahim Karimi}
\email{ekarimi@uottawa.ca}
\affiliation{The Max Planck Centre for Extreme and Quantum Photonics, Department of Physics, University of Ottawa, 25 Templeton St., Ottawa, Ontario, K1N 6N5 Canada}
\affiliation{Department of Physics, Institute for Advanced Studies in Basic Sciences, 45137-66731 Zanjan, Iran}
%
\begin{abstract}
Free electrons with a helical phasefront, referred to as \qo{twisted} electrons, possess an orbital angular momentum (OAM) and hence a quantized magnetic dipole moment along their propagation direction. This intrinsic magnetic moment can be used to probe material properties. Twisted electrons thus have numerous potential applications in materials science. Measuring this quantity often relies on a series of projective measurements that subsequently change the OAM carried by the electrons. In this Letter, we propose a non-destructive way of measuring an electron beam's OAM through the interaction of this associated magnetic dipole with a conductive loop. Such an interaction results in the generation of induced currents within the loop, which are found to be directly proportional to the electron's OAM value. Moreover, the electron experiences no OAM variations and only minimal energy losses upon the measurement, and hence the non-destructive nature of the proposed technique.
\end{abstract}
\pacs{Valid PACS appear here}
\maketitle

\noindent Electrons can possess net quantized orbital angular momentum (OAM) while undergoing free-space propagation~\cite{bliokh:07}. The wavefunction $\psi$ associated with such an electron includes an $\exp{(i\ell\varphi)}$ term arising from its helical phase-fronts, where $\ell$ and $\varphi$ are an integer and the azimuthal coordinate, respectively. Beams consisting of these \qo{twisted} electrons are referred to as electron vortex-beams. Different techniques, such as direct imprinting of a phase variation~\cite{uchida:10}, amplitude~\cite{verbeeck:10} and phase~\cite{grillo:14} holograms, and magnetic needles~\cite{beche:14} have experimentally been shown to generate such electron beams. In turn, these electron beams possess quantized OAM and circulating current densities $J_\varphi$ in a plane orthogonal to their propagation direction. It thus follows that these current densities cause twisted electron beams to carry a magnetic dipole moment $\ell\, \mu_\text{B}$ in addition to their intrinsic spin magnetic dipole moment $\pm\,\mu_\text{B}$, where $\mu_\text{B}$ is the Bohr magneton~\cite{bliokh:12}. Hence, unlike its intrinsic spin, the magnetic moment associated with its twisted wavefront is in principle unbounded, allowing  values as high as $200\,\mu_\text{B}$ to be achieved experimentally~\cite{mcmorran:11, grillo:15}. Such a large unbounded magnetic moment may find applications in materials science~\cite{harris:15}, overcoming the fact that the generation of spin-polarized electron beams has historically been affected by empirical and fundamental difficulties~\cite{karimi:12}. Among future potential applications are investigations related to magnetic dichroism in materials~\cite{lloyd:12}, the fundamental nature of radiation~\cite{ivanov:13}, exotic physics such as virtual forces~\cite{kaminer:15}, and the interaction of twisted electrons with light beams~\cite{hayrapetyan:14}. Many of these examples require the analysis of the electron beam's OAM content, a process adopted from its optical counterparts and that is usually carried out by making the beam go through phase-flattening projective measurements by means of phase holograms~\cite{mair:01,qassim:14,saitoh:13}. However, the analysis of each OAM component requires the use of a distinct hologram, that can make the investigation of a beam's OAM components long, tedious, and inefficient. Moreover, the beam's OAM content, after passing through a phase mask, will have a value different from that of the initial state~\cite{qassim:14}. 

In this Letter, we propose an alternative way of measuring an electron beam's OAM relying on electric fields induced by time-varying magnetic fields. The principle of our technique is related to one where a magnet is dropped through a conductive tube (or ring). The falling motion of the magnet generates currents within the tube, that in turn produce a magnetic force countering the magnet's descent~\cite{saslow:92,maclatchy:93,donoso:11}. By using a similar reasoning, in the non-relativistic regime, one can calculate the induced current inside a micro-scale conductive ring due to the motion of an OAM-carrying electron traveling through it. Because the electron's OAM and magnetic moment are quantized, the magnetic field emanating from the electron will also be quantized and will produce discrete induced currents inside the ring that can be related directly to the OAM carried by the electron.

\noindent We use a semi-classical approach to describe the interaction between a propagating electron vortex beam and a conductive material. Let us consider an electron with a rest mass $m_e$ propagating along a specific axis, e.g. the $z$-axis, and possessing a well-defined central kinetic energy ${\cal E}$ and momentum $p_0$, where $c$ is the velocity of light in vacuum. Under the slowly-varying amplitude approximation, the wavepacket associated with this electron must satisfy the paraxial Schr\"odinger equation. The corresponding wavefunction is quantized, and holds a specific shape based on its initial probability and phase distribution conditions. For instance, it may be quantized in the transverse plane as well as in the longitudinal direction~\cite{bliokh:07}, which yields the following wavepacket in cylindrical coordinates $r,\varphi,z$
\begin{align}\label{eq:wavefunction}
	\psi_{p,\ell,n}(r,\varphi,z;t)=u_{p,\ell}^\text{LG}(r,\varphi;t)\,u_{n}^\text{HG}(\zeta)\,e^{i\left(p_0 z-{\cal E} t\right)/\hbar},
\end{align}
where $u^\text{LG}$ and $u^\text{HG}$ are Laguerre-Gauss and Hermite-Gauss modes~\cite{siegman:86}, respectively, in which $p$ and $n$ are positive integers defining the electron's distribution in the transverse plane and the longitudinal direction. $\ell$ is an integer number that is associated with the OAM carried by the beam and also defines its transverse distribution. The electron wavepacket's centre of mass is denoted by $\zeta=z-p_0 t/m_e$ while $\hbar$ is the reduced Planck constant. On account of the electron's OAM, its rest frame four-current density consists only of a scalar and an azimuthal component, according to the expression 
\begin{align}\label{eq:four-current}
	\mathbf{j}_\text{rest}^{\alpha}=\left(c\rho, J_r, J_\varphi,J_z \right)=\left(-c e\,{\cal P}, 0, \frac{\hbar \ell}{m_e r}\,{\cal P}, 0\right),
\end{align}
where $\rho$, $J_r$, $J_\varphi$, and $J_z$ correspond to charge density, and radial, azimuthal, and longitudinal current densities, respectively; while ${\cal P}={\cal P}(r',\varphi';z')=|\psi_{p,\ell,n}(r,\varphi,z;t)|^2$ is the probability density function of the electron's position in its rest frame defined by the coordinates $r',\varphi',z'$; and $-e$ is the electron charge. The four-current densities in the laboratory frame that the electron perceives as traveling along the $z$-direction can then be calculated via an inverse Lorentz transformation $\mathbf{j}_\text{lab}^{\alpha}=({\Lambda^{\alpha}_{\beta}})^{-1}\,\mathbf{j}_\text{rest}^{\beta}$, yielding $\mathbf{j}_\text{lab}^{\alpha}=\left(-c e\,\gamma\,{\cal P}, 0, \frac{\hbar \ell}{m_e r}\,{\cal P}, -\gamma\,\beta\,c \, e\,{\cal P}\right)$, where $\Lambda^{\alpha}_{\beta}$ is the Lorentz transformation matrix, $\beta=p_0/(m_e\,c)$ and $\gamma=(1-\beta^2)^{-1/2}$~\cite{jackson:99}. Likewise, a Lorentz boost along the $z$-axis must also be applied to the electron's rest-frame coordinates to express its current densities with respect to the laboratory frame coordinates, i.e., $\mathbf{x}_\text{lab}^{\alpha}=({\Lambda^{\alpha}_{\beta}})^{-1}\,\mathbf{x}_\text{rest}^{\beta}$, where $\mathbf{x}^{\alpha}=(c\,t,\mathbf{r})$. One may associate the first, third, and last terms of the four-vector current density with an electrostatic potential $\cal V$, and the azimuthal and longitudinal vector potentials $\mathbf{A}_{\varphi}$ and $\mathbf{A}_{\parallel}$, respectively.  The azimuthal current density $J_\varphi=({\hbar \ell}/{m_e r})\,{\cal P}\,\mathbf{e_\varphi}$ generates a magnetic field $\mathbf{B}=\nabla \times\mathbf{A}$ oriented along the electron's propagation direction, i.e. the $z$-axis, where $\nabla$ is the gradient operator, and $\mathbf{e_\varphi}$ is the azimuthal unit vector. The vector potential $\mathbf{A}_{\varphi}$ at a given position $\mathbf{r}$ can then be expressed directly as a solution to one of Poisson's equations, namely $\mathbf{A}_\varphi(\mathbf r)=\mu_0/(4\pi)\int d^3\mathbf{r'}\, G(\mathbf{r},\mathbf{r}')\,{J_\varphi(\mathbf{r}')}$, where $G(\mathbf{r},\mathbf{r}')=|\mathbf{r}-\mathbf{r}'|^{-1}$ is the corresponding Green's function. The electron's transverse motion for any value of $\ell$ is then considered as a \qo{localized} current loop defined by $I_e=e\hbar/(\pi\,m_e w_0^2 )\,\mathbf{e}_\varphi$, as prescribed by the relation $\ell \mu_\text{B}=I_e (\pi r_\ell^2)$, where $r_\ell=w_0\sqrt{\ell/2}$ is the radius at which an electron is maximally distributed and $w_0$ is the minimum radius of its Gaussian distribution.

The vector potential associated with such an azimuthal current can be expressed in the form
\begin{align}\label{eq:vector}
	\mathbf{A}_\varphi(r,z)=\frac{\mu_0 I_e\,\eta}{\pi v^{3/2}}\,\left(u\,K\left(2\eta^2\right)-(u+v) E\left(2\eta^2\right)\right),
\end{align}
where $u=r_\ell^2+r^2+z^2$, $v=2r_\ell r$, $\eta=v^{1/2}(u+v)^{-1/2}$, and $K(.)$ and $E(.)$ are the complete elliptic integrals of the first and second kind, respectively~\cite{jackson:99}. 
\begin{figure}[h]
	\begin{center}
	\includegraphics[width=0.9\columnwidth]{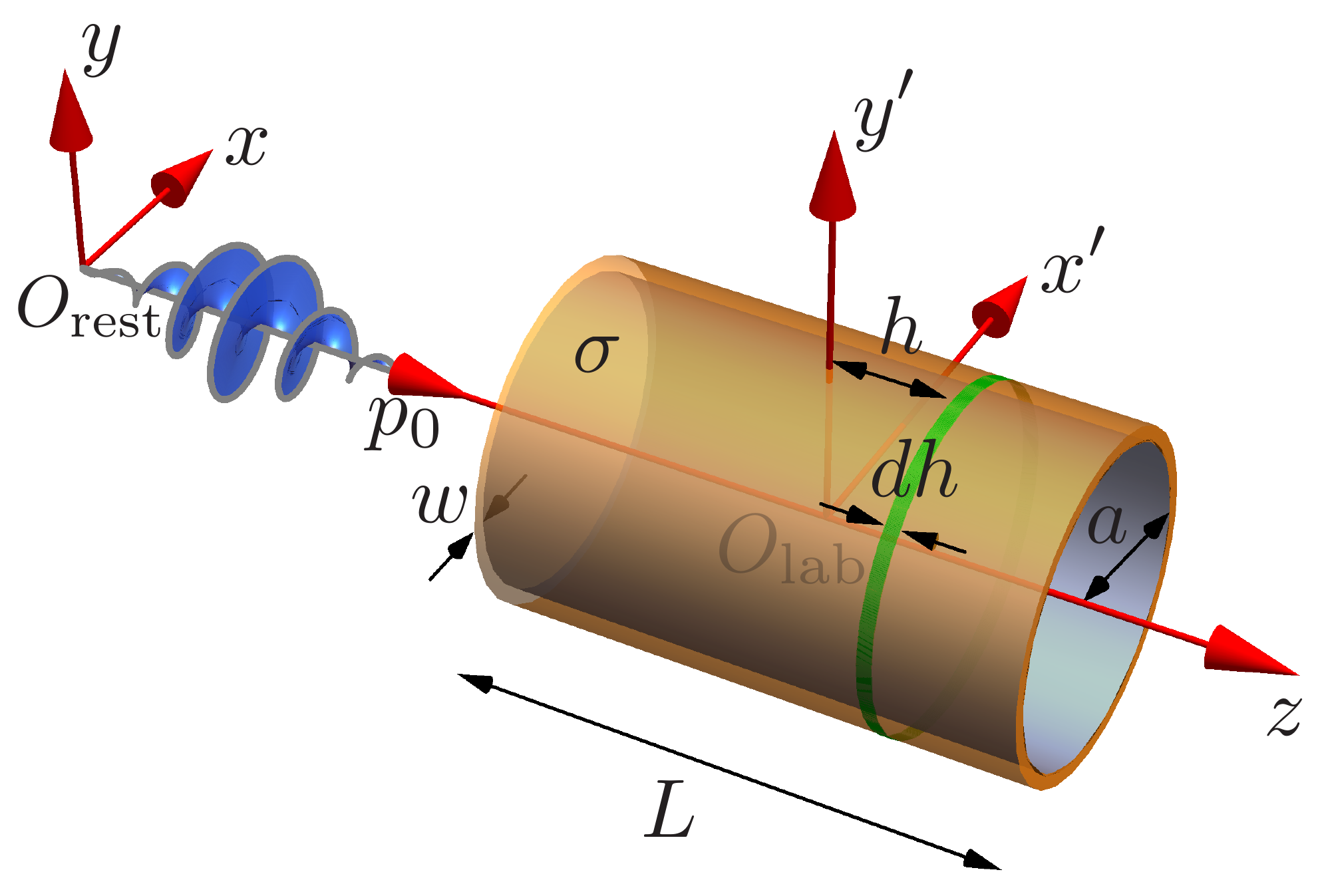}
	\caption[]{System in which an electron vortex beam with a central energy $\cal E$ and momentum $p_0$, which is in its lower longitudinal mode ($u_{p,\ell}^\text{LG}(r,\varphi;t)\,u_{0}^\text{HG}(\zeta)\,e^{i\left(p_0 z-{\cal E} t\right)/\hbar}$), propagates through a cylinder with conductivity $\sigma$ and permeability $\mu$. The relative motion of both entities results in the generation of a current in the infinitessimal loop of thickness $dh$.}
	\label{fig:fig1}
	\end{center}
\end{figure}
As depicted in Fig.~\ref{fig:fig1}, we consider such electrons passing through a tube of thickness $w$, radius $a$, conductivity $\sigma$ and length ${L}$. The tube radius is large enough to ensure that the electron's wavefunction nearly vanishes at its inner radius. In particular, for $p=0$ mode distributions defined by an arbitrary $\ell$ index, the tube radius $a$ is chosen to be much greater than the radius $r_\ell$, i.e. $a\gg r_\ell$. The conductive tube can be considered as a sequence of infinitesimal circle-loops positioned at a longitudinal distance $h$ from the tube's centre. As predicted by Faraday's law of induction, when the twisted electron travels through the tube, its longitudinal magnetic field induces an eddy current in each of the tube's infinitesimal loops. According to Lenz's law, the direction of these currents must generate a magnetic field that is opposed to the motion of the electron beam. Its value, however, will depend on the time variation of the magnetic flux $\Phi_\text{B}$ through each loop, i.e. $-\partial_t \Phi_\text{B}$. Neither the electrostatic potential $\cal V$ nor the longitudinal vector potential $\mathbf{A}_{\parallel}$ contributes to the magnetic flux $\Phi_\text{B}$. Only the azimuthal vector potential $\mathbf{A}_{\varphi}$ is relevant to the analysis. Due to the cylindrical symmetry of the electron-tube system, the vector potential is also independent of $\phi$. 
The induced electric field on the circle-loop located at position $h$, and hence the induced current, is therefore azimuthal and expressed as $E_\varphi=-\partial_t \mathbf{A}_\varphi(a,z-h)$ where $z$ is the electron's relative longitudinal position. One can show, by means of Ohm's law, $dI=\sigma E_\varphi\,(w\,dh)$, that the total current within the tube induced by an electron with a magnetic dipole moment $\ell\mu_B$ is given by the expression
\begin{align}
	\label{eq:current}
	I = \frac{3}{4\pi}\left(\frac{p_0}{m_e}\right)\,(\sigma\mu\,w a)\,\left(\ell \mu_B\right)\int_{-L/2}^{L/2}\frac{\gamma^2(z-h)\, dh}{(a^2+\gamma^2(z-h)^2)^{5/2}}\,,
\end{align}
where $\mu$ is the tube's permeability. The proportionality of this relation describes the quantization of the induced current within the tube due to the discrete nature of the electron's OAM. By integrating Eq.~({\ref{eq:current}}), an analytical expression for this current can be obtained and is plotted as a function of electron position relative to the cylinder's centre in Fig.~\ref{fig:fig2}-(a) for various values of electron OAM. As a result, one can conceive a device for OAM measurement by detecting the corresponding quantized current induced inside a tube or a thin loop circuit.
\begin{figure}[h!]
	\includegraphics[width=1\columnwidth,left]{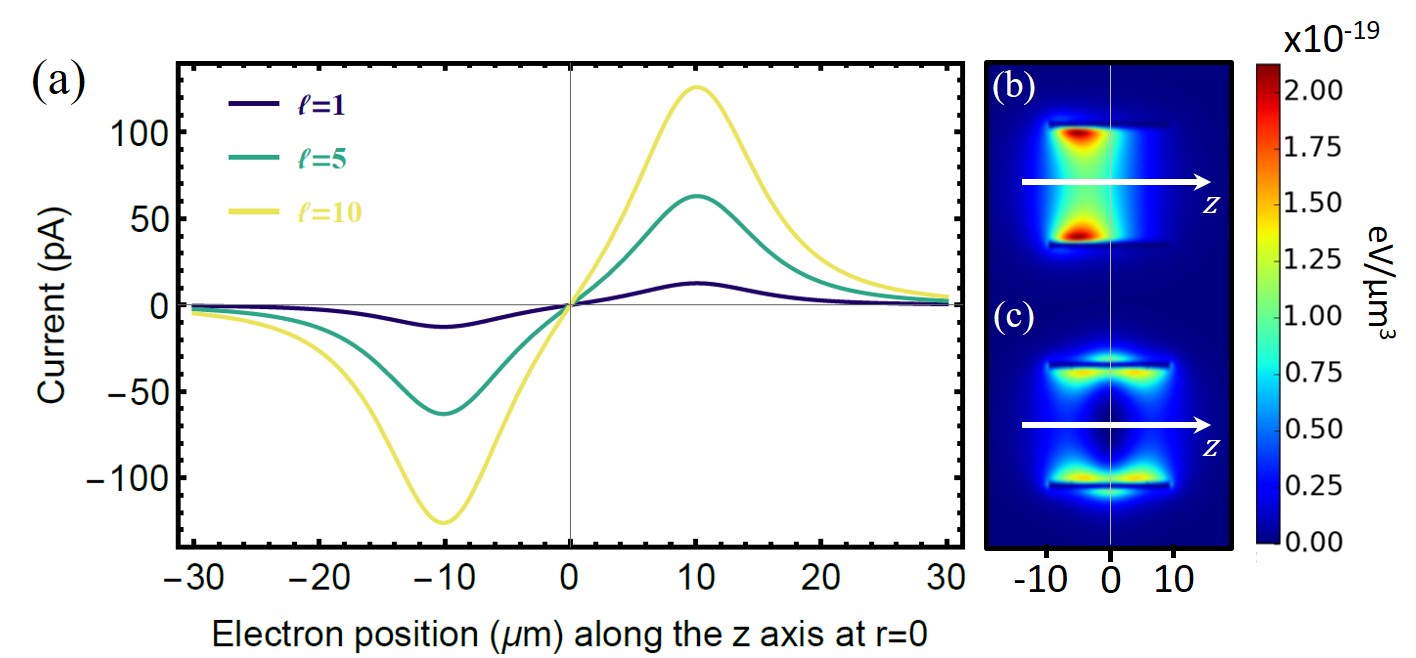}
	\caption[]{(a) Theoretically calculated total induced current in a conductive tube by an electron vortex beam. We assumed that the electron beam carries OAM of $\ell=1,5$ and $10$, and that the conductive tube is made of platinum. Longitudinal cross section of the tube depicting the relative magnetic energy density generated by its induced eddy currents when an electron consisting of a high OAM quantum ($\ell=100$) is (b) entering the tube and (c) in the middle of the tube. Here, we assumed an electron beam with central energy ${\cal E}=100\,keV$, and a platinum tube with length $L=20\,\mu m$, thickness $w=1\,\mu m$ and radius $a=10\,\mu m$.}
	\label{fig:fig2}
\end{figure}
As shown in Fig.~\ref{fig:fig2}-(a), currents of order of tens of $pA$ are induced in the loop and could be potentially read out using an ampere meter (e.g. Tektronix 6485 Picoammeter). Therefore, this technique can potentially be used to measure OAM values of twisted electron beams. The direction of the induced current additionally provides information on the sign of the OAM value. Moreover, since the generated current is directly proportional to the material's conductivity, it follows that by using a more conductive material one could increase the current by several orders of magnitude. Though these induced currents are rather short-lived, a combination of fast electronics, optimized cylinder dimensions, and secondary methods such as autocorrelation techniques can be used to overcome experimental difficulties related to the short interaction between the electron and the cylinder~\cite{suppmat}. 
Our proposed technique has no influence on the OAM of the electron beam since the electron's canonical OAM is conserved in the presence of an external longitudinal magnetic field~\cite{Barnett:14}. The only property that the measurement affects is the energy carried by the electron~\cite{note:01}. This is due to the fact that the induced currents will counter the motion of the electron. The energy loss due to the electron-tube interaction is $\Delta {\cal E}=-(2\pi \sigma a w)\,\left(\frac{p_0}{m_e}\right)\,\int_{-\infty}^{+\infty}\,dz\int_{-L/2}^{L/2}\left(\partial_z\mathbf{A}_\varphi(a,z+h)\right)^2\,dh$, which slightly decelerates the electron. This deceleration can potentially reach a relatively high value, resulting in a large radiated electromagnetic power emitted from the electron, as indicated by the Larmor formula. However, due to its very short time of interaction with the tube, the energy lost by the electron can only realistically reach a value on the order of $10^{-20}$ eV when an electron with an OAM $\ell=100$ is considered (we assume the parameters reported in Fig.~\ref{fig:fig2}). Such energy values are obtained when deducing the force applied on the electron by the tube. Another way of obtaining insight on the electron's energy loss is to calculate the total energy contained within the fields generated by the relative motion of the electron. This energy can be calculated numerically by first finding the magnetic field generated by the cylinder's loops of currents using the Biot-Savart law~\cite{good:01} and then integrating the total energy stored in these magnetic fields and within the electric fields associated with the currents themselves. In particular, this method was employed to produce the energy density plots found in Fig.~\ref{fig:fig2} (b)-(c). One can see that the act of measuring the electron's OAM has nearly no effect on the electron itself. Indeed, unlike projective measurement techniques, where the electron's phase-front is flattened and projected on a Gaussian mode, the electron's OAM does not change during the measurement. For these reasons, the electron's motion through the tube leaves it largely unperturbed. Up until this point, we only considered the case where an electron travels perfectly along the center of the tube. A simple extension of this analysis reveals that the calculated induced currents are not significantly affected by breaking the apparatus' cylindrical symmetry. We further discuss how the currents are affected by asymmetries in experimental apparati in the supplemental material~\cite{suppmat}.

\begin{figure}[h!]
	\begin{center}
	\includegraphics[width=0.92\columnwidth]{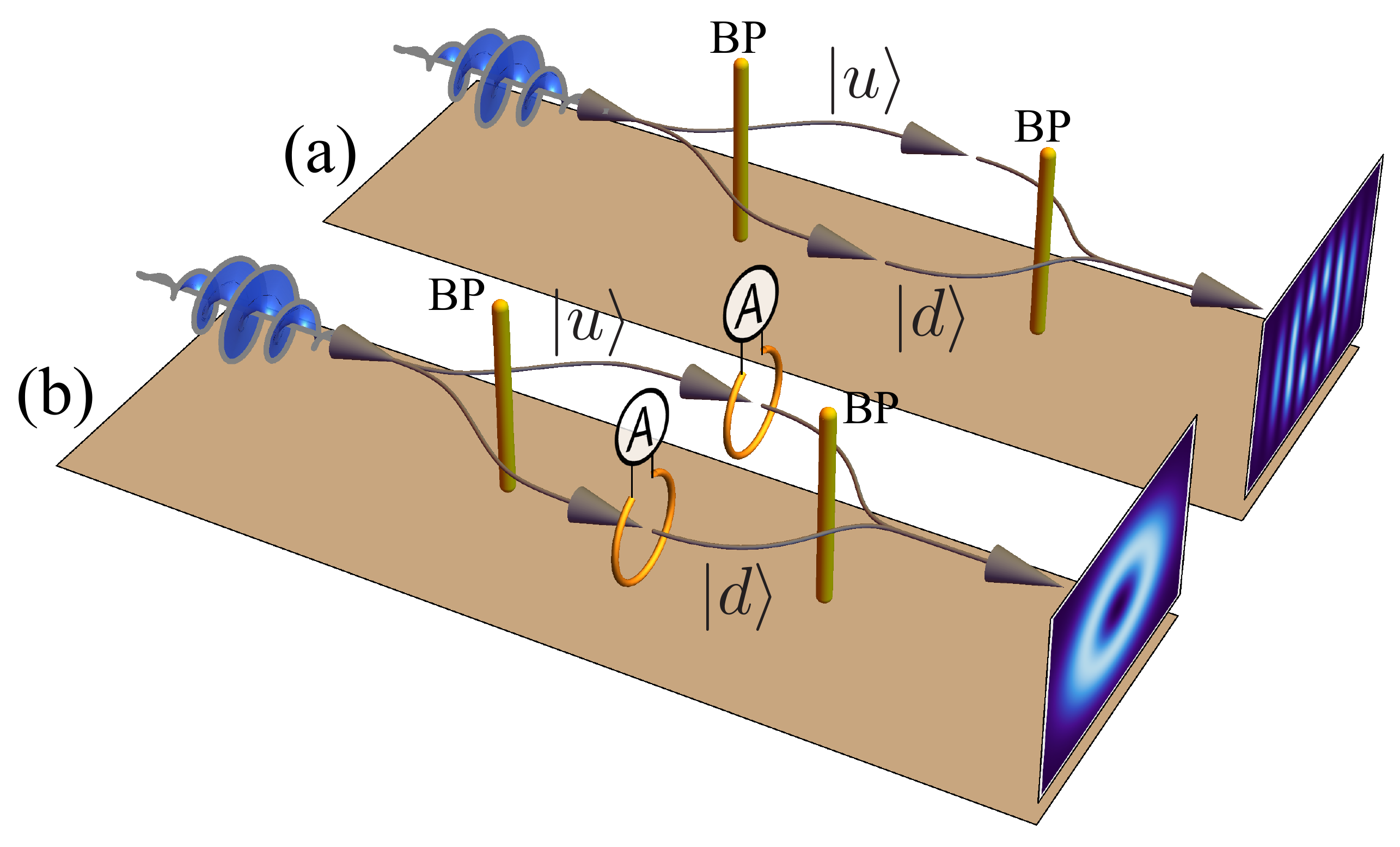}
	\caption[]{Proposed experiment in which the effect of conductive circuits in an OAM-carrying electron double slit experiment is considered. (a) The electron double slit experiment in which no circuits are present. (b) The electron double-slit experiment in which a circuit is present in each of the possible paths, $\ket{u}$ or $\ket{d}$, taken by the electron where the coefficent $\alpha=0$ (i.e. the electron and the loops are perfectly coupled). BP annotations refer to biprisms.}
	\label{fig:fig3}
	\end{center}
\end{figure}

This non destructive approach to measuring OAM may create a conceptual paradox. One may mistakenly argue that because this measurement leaves the electron's quantum state (OAM and energy) unchanged, it could challenge the validity of a wave-particle duality experiment (quantum complementarity). Consider a double-slit experiment in which, due to an electrostatic interaction, the electron wavefunction is split into two parts $\ket{u}$ and $\ket{d}$. Both parts are then coherently recombined and interfere at a screen, as illustrated in Fig.~\ref{fig:fig3}(a)~\cite{uchida:13}. The electron's state can be described as a superposition of both paths, as if it is in a coherent superposition of states $\ket{u}$ and $\ket{d}$, resulting in the formation of an interference pattern on the screen. In the case when the electron is equally likely to take each path, its state may be described as $\ket{\psi}=(\ket{u}+e^{i\delta}\ket{d})/\sqrt{2}$ where $\delta$ is the relative phase between the states. The corresponding density matrix is pure $\rho:=\ket{\psi}\bra{\psi}=(\ket{u}\bra{u}+e^{-i\delta}\ket{u}\bra{d}+e^{i\delta}\ket{d}\bra{u}+\ket{d}\bra{d})/2$. In this expression, the terms $e^{-i\delta}\ket{u}\bra{d}$ and $e^{i\delta}\ket{d}\bra{u}$ carry the interference pattern's phase information and can therefore be associated with the fringe visibility, which is unity for this ideal case. The terms $\ket{u}\bra{u}$ and $\ket{d}\bra{d}$, respectively, describe the probability of finding an electron in the $\ket{u}$ or $\ket{d}$ path, both of which are equiprobable events for this case~\cite{englert:96}.

Now, consider two conductive circuits introduced into each of the possible paths, as shown in Fig.~\ref{fig:fig3}(b). As mentioned above, these circuits have the capacity to measure an electron's OAM with minimal energy loss, allowing for the detection of whether an OAM-carrying electron has taken a given path. When no electron travels through the circuit, the circuit is in a state $\ket{0}_c$. When an electron with an OAM number $\ell$ travels through the circuit, it will induce a quantized current, changing the circuit to a state defined by $\ket{i^\ell}_c$ which can be expressed as a superposition of the loop's current eigenstates $\ket{n}_c$, i.e. $\ket{i^\ell}_c = \sum_{n}c_n\ket{n}_c$, where $c_n=\braket{n}{i^\ell}_c$ is an expansion coefficient depending on various experimental parameters describing the interaction between the free electron and the loop itself. Hence, the system consisting of both circuits can be described by the tensor product $\ket{i_u}_c\ket{i_d}_c$ where $\ket{i_u}_c$ and $\ket{i_d}_c$, respectively, represent the state associated with the electrical current going through circuits in the $\ket{u}$ and $\ket{d}$ paths.
Because the circuit has the ability to provide information about which path the electron has taken, it provides information about the particle nature of the electron, while the presence of fringes gives information about its wave nature. Therefore, the presence of currents and fringes would seemingly allow one to detect both the electron's wave and particle nature simultaneously, violating the principle of complementarity.

However, prior to going through any of the circuits, the system consisting of the electron and the two circuits can initially be expressed as $\ket{\psi_i}=\tfrac{1}{\sqrt{2}}(\ket{u}+e^{i\delta}\ket{d})\ket{0_u}_c\ket{0_d}_c$. After the electron has gone through either of the circuits, the system's final wavefunction becomes $\ket{\psi_f}=\tfrac{1}{\sqrt{2}}(\ket{u}\ket{i^\ell_u}_c\ket{0_d}_c+e^{i\delta}\ket{d}\ket{0_u}_c\ket{i^\ell_d}_c)$, where the electron is entangled with the circuits. The circuits thus act as a nonlocal \qo{environment} and cause the electron state to partially decohere~\cite{zurek:03}. In order to observe the effect of the circuits' presence on the obtained interference pattern, we take the partial trace over the circuits' states. The reduced density matrix will correspond to $(\ket{u}\bra{u}+e^{-i\delta}\alpha\ket{u}\bra{d}+e^{i\delta}\alpha^*\ket{d}\bra{u}+\ket{d}\bra{d})/2$, where $\alpha=\braket{0_u}{i^\ell_u}\braket{i^\ell_d}{0_d}$. One can observe that the visibility terms of the reduced density matrix in the $\{ \ket{u},\ket{d}\}$ basis will be modified by the factor $\alpha<1$, where for identically coupled circuits, i.e. $\ket{i^\ell_d}=\ket{i^\ell_u}$, $\alpha=\abs{c_0}^2$. This coefficient, defined by $\braket{0}{i^\ell}_c$, will vary with the coupling between the free electron and the circuit's state which is determined by various experimental parameters. Such parameters, which include the circuit's radius for instance, can be modified to provide a varying $\alpha$ coefficient affecting the fringe visibility.

In conclusion, we present a non-destructive technique that can be used to measure the OAM of an electron beam. The technique is based on the interaction of the quantized magnetic dipole moment of the twisted electron and a conductive tube. The beam's OAM components are measured by detecting the quantized induced eddy currents in the tube. These electrons suffer minimal energy losses and the method is non-destructive. To illustrate the limitations of the method, we also describe the possibility of using such a device in a gedanken quantum experiment, in which the knowledge of an electron's presence is needed. Doing so would result in reducing the visibility of observed interference as prescribed by complementarity. 
A prospective extension to the method could be using the tube to generate radiation with an approach similar to that of ~\cite{mohammadi:12} through the formation of plasmons by introducing a discontinuity in the tube, such as the absence of conductive material at a given azimuthal angle. However, this would result in larger energies being lost by passing electrons. This method's minimal electron energy loss is an essential aspect to its non-destructive nature which, along with the preservation of the electron's original OAM, presents this technique as a viable alternative to modern projective measurements.

\textit{Acknowledgments:} E.K. thanks Profs. Gerd Leuchs and Israel De Leon for fruitful discussions involving the topic. H.L., F.B, A.S., R.W.B. and E.K. acknowledge the support of the Canada Research Chairs (CRC) and Canada Excellence Research Chairs (CERC) Program. R.D.B. thanks the European Research Council for an Advanced Grant.
%

\bigskip
\section*{Supplemental material for ``Non-destructive measurement of an electron beam's orbital angular momentum''}
\footnotesize
\subsection{Experimental Schematics}
\begin{figure}[h]
	\begin{center}
	\includegraphics[width=\columnwidth]{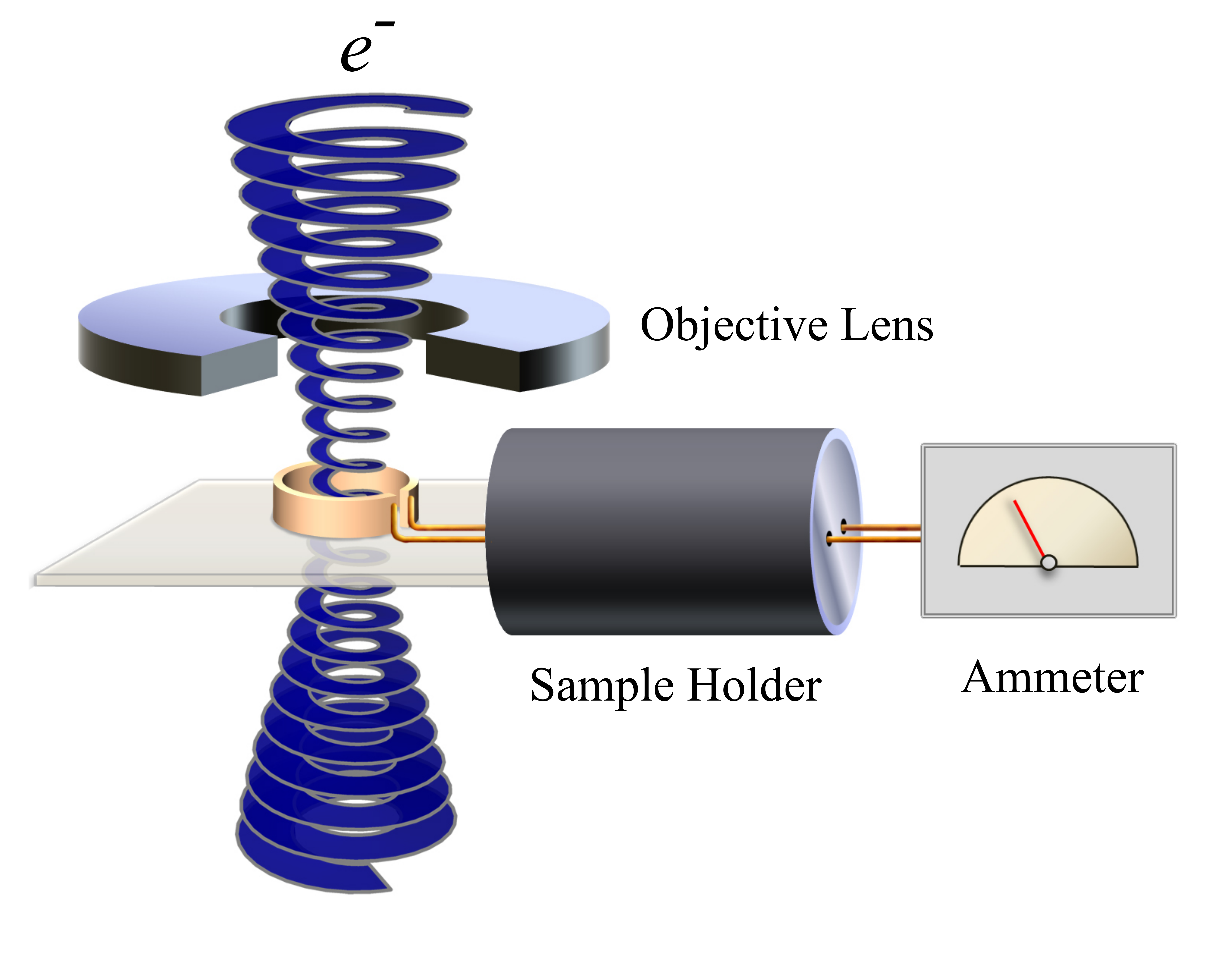}
	\caption[]{Proposed experimental schematic for the non-destructive measurement of an electron beam's orbital angular momentum. An OAM-carrying electron beam, which possesses quantized magnetic dipoles, is being focused through a conductive loop. The magnetic dipole induces a quantized current in the loop that is read out by an ammeter.}
	\label{fig:schematics}
	\end{center}
\end{figure}

We propose the following electron microscope-based apparatus, whose schematic is provided in Fig.~\ref{fig:schematics}, to non-destructively measure an electron's orbital angular momentum (OAM). The OAM-carrying electron beam passes through an objective lens focusing it through the conductive loop. The latter is held by the microscope's sample holder and also possesses a thin slit in which leads are placed and connected to an ammeter. As the electron passes through the conductive loop, it induces quantized Eddy currents associated with its OAM, given by Eq. (4) of the main text, which are thereafter read out by the ammeter. 
Should the discontinuity introduced by the thin slit provide any obstacles hindering our proposed measurement, an alternative way of measuring a passing electron's OAM would be to wrap the cylinder with a nanoscale solenoid. Much like the cylinder itself, the solenoid would pickup the electromotive force caused by the passage of the electron, thus resulting in the generation of quantized currents within its coils.

\subsection{Asymmetrical Considerations}
\begin{figure}[h!]
	\begin{center}
	\includegraphics[width=0.9\columnwidth]{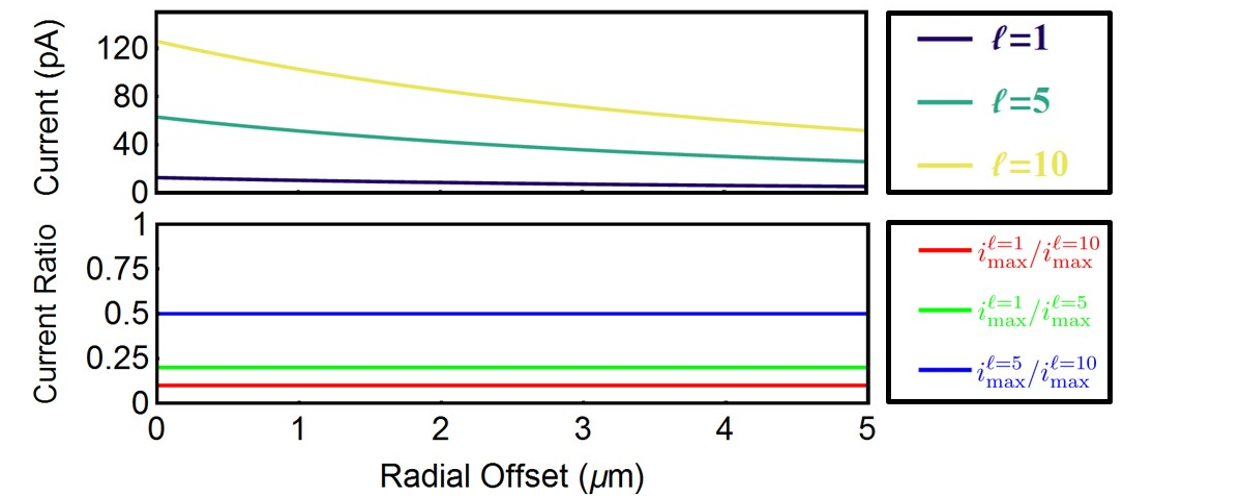}
	\caption[]{Maximum induced current induced by the electron upon its propagation as a function of the electron's radial offset along with the corresponding relative ratio between currents induced by electrons defined by different OAM values. These currents correspond to those in the region of the cylinder that will experience the greatest drop in electromotive force upon breaking its symmetry.}
	\label{fig:offset}
	\end{center}
\end{figure}

We now consider how the induced currents are affected by breaking the apparatus' cylindrical symmetry. We assume that the most likely imperfection is that the electron trajectory is offset in the transverse direction from its ideal position on the axis of the cylinder. Placing an interacting element perpendicularly to an electron beam is usually not a practical issue. As illustrated in Fig.~\ref{fig:offset}, where we considered the parameters employed in Fig.~2 of the main text, a simple
extension of our analysis reveals that the order of magnitude of the currents related to the electron's OAM is not affected by these offsets.
In particular, the currents shown in Fig.~\ref{fig:offset} are found based on the azimuthal electric field induced in the region of the conductive loop that will experience the lowest induction. The ratio between such currents defined by passing electrons with different OAM values remains constant. Therefore, breaking the apparatus' cylindrical symmetry will not affect the detectable nature of the electrons nor will it change the relative responses associated with electrons of different OAM values.

Though the electron's electric field may also add unwanted contributions to these currents, they do not vary with the electron's OAM. These unwanted contributions may also be suppressed by optimizing certain experimental parameters such as the device's transverse position with respect to the axis along which the electron is displaced.

\subsection{Suggested Measurement Methods}
A potential obstacle to our proposed OAM measurement consists of the rather short interaction between the conductive loop and the electron. In particular, for the parameters used in Fig.~2 of the main text, the duration of this interaction is within the order of 0.1~ps, thus making the realizability of this measurement particularly challenging. 
Though the frequencies of such induced currents can potentially reach high values, e.g. in the order of $10^{13}$ Hz, the fact that they are significantly below those of optical frequencies refrains the occurence of anomalous phenomena such as the generation of surface plasmons. Therefore, the only obstacle that this short interaction introduces is the challenge of measuring it.
However, there are some electronics with ps resolution (e.g. Hydraharp) that can make such a task much easier. Moreover, because the electron's time of passage through the loop scales with the loop's length, a longer cylinder can be used to make the interaction more noticeable. Likewise, lower energy electrons can also be employed to achieve the same effect. Such elementary measures can, to a certain extent, overcome difficulties related to this short interaction. We also propose two additional methods to address issues associated with the electrons's time of passage through the loop.

\begin{figure}[!b]
	\begin{center}
	\includegraphics[width=0.9\columnwidth]{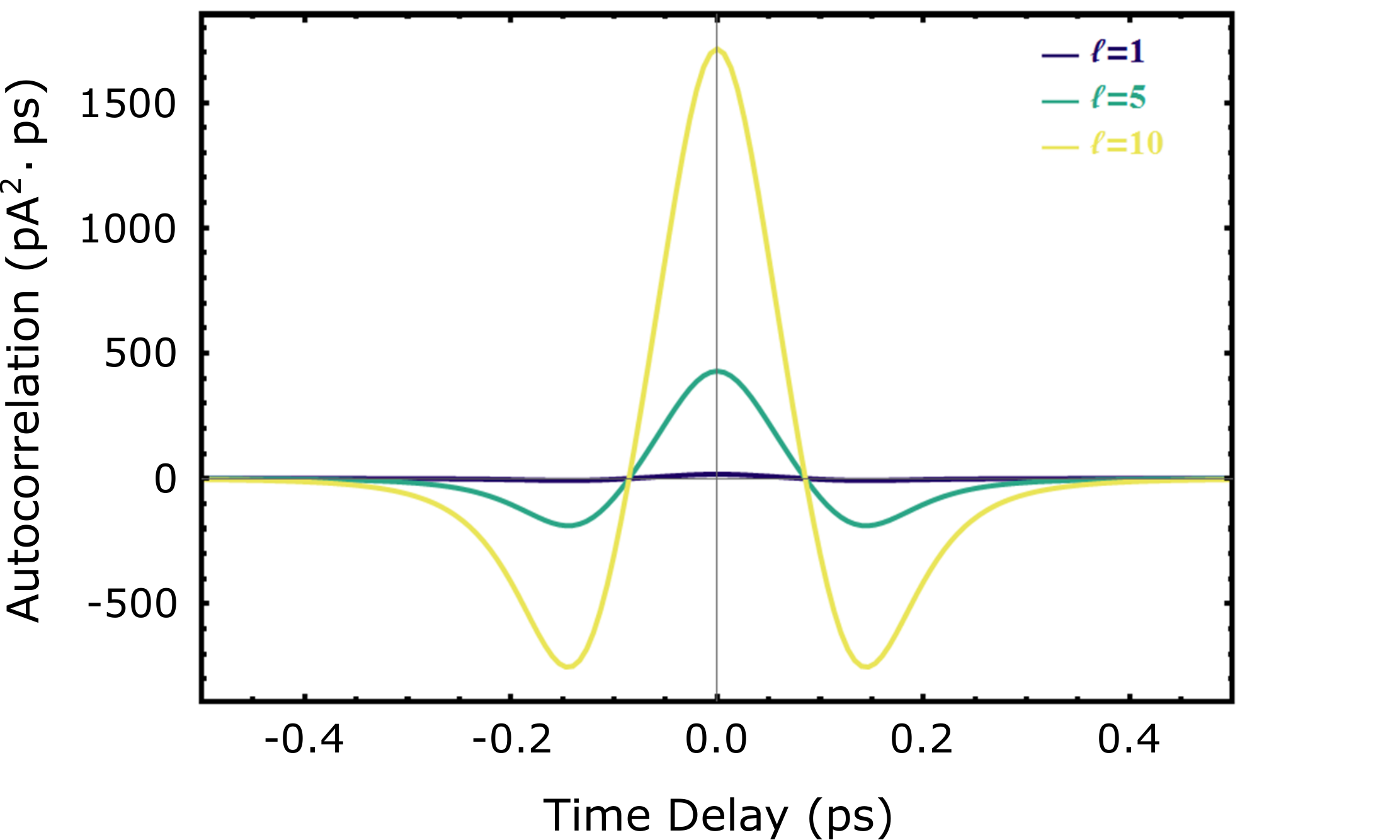}
	\caption[]{Autocorrelation functions associated with electrons carrying OAM values of $\ell=1$, 5, 10. The functions were calculated using the electron and loop parameters employed in Fig.~2 of the main text.}
	\label{fig:autocorrelation}
	\end{center}
\end{figure}

On one hand, the electron's OAM can be measured with an autocorrelation method involving the use of two conductive loops in which currents will be induced by the passage of distinct electrons. With these currents, one can calculate the electron beam's autocorrelation trace, $C(\tau)$, as a function of the time delay between the electrons' passage through their respective loops. This autocorrelation function is provided in Eq.~(\ref{eq:autocorrelation})
\begin{align}
	\label{eq:autocorrelation}
	C(\tau) \propto \int_{-\infty}^{\infty} i^\ell(t)i^\ell(t-\tau)\,dt.
\end{align}
This trace will scale with the square of the electron's OAM as depicted in Fig.~\ref{fig:autocorrelation} where we provide the autocorrelation functions calculated using the parameters employed in Fig.~2 of the main text.

\begin{figure}[t]
	\begin{center}
	\includegraphics[width=0.9\columnwidth]{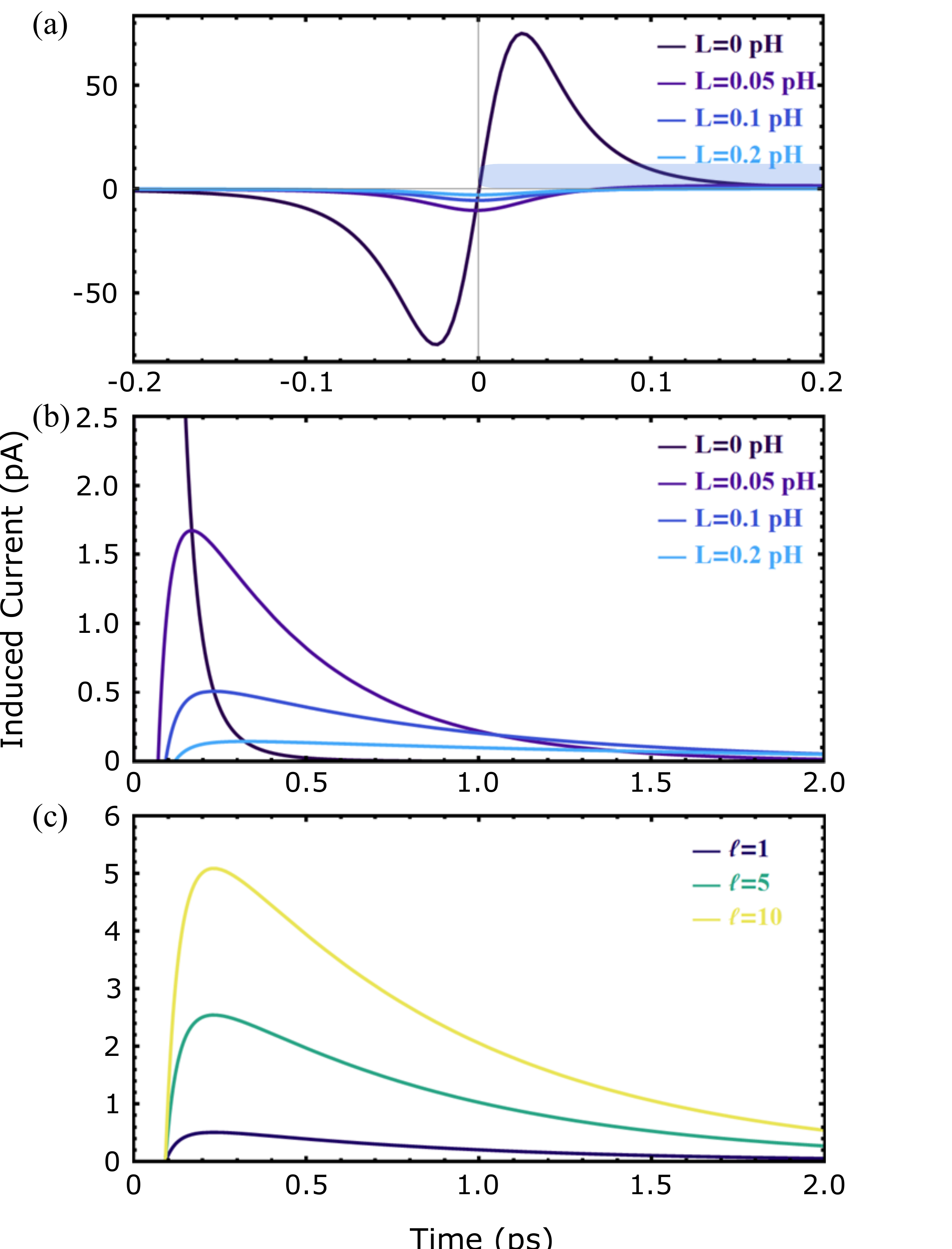}
	\caption[]{(a) Currents induced in circuits with inductances of $0$, $0.05$, $0.1$, and $0.2$ pH by an electron carrying one unit of OAM. (b) Zoomed in region of (a) over an extended time scale. (c) Currents induced in a circuit with an inductance of  $0.1$ pH by electrons carrying OAM values of $\ell=1$, $5$, and $10$.}
	\label{fig:inductionfig}
	\end{center}
\end{figure}

On the other hand, methods to ``extend'' the lifetime of the induced currents would also make them more noticeable. For instance, an inductor could be added to the circuit consisting of the ammeter and the conductive loop. According to Lenz's law, the rate at which the measured current increases as the electron enters the loop will significantly be decreased. However, as the electron exits the loop, the circuit's current will be maintained by the inductor in order to minimize the rate at which the magnetic flux through it decreases. Therefore, though an inductor will lower the registered currents to smaller yet still detectable magnitudes, it will also extend their duration.

We illustrate this concept in the following. First, we assume that the electromotive force induced by the electron's passage inside the loop is not affected by the addition of the inductor. Second, to simplify, we assume that the EMF induced inside the loop is uniform over its length as opposed to what was calculated in the main text. Last, we assume that our system consisting of the loop and the inductor can be represented by a series RL circuit in which the driving voltage is the electromotive force induced inside the cylinder $\mathcal{E}(t)$, the resistance is that associated with the dimensions and the material of the cylinder $R$, and the inductance is that of the inductor $L$. The resulting differential equation describing the induced current's $I(t)$ time evolution is thus
\begin{align}
	\label{eq:inductor}
	\mathcal{E}(t)-L\frac{dI}{dt}=I\,R.
\end{align}
We numerically solve Eq.~(\ref{eq:inductor}) for the parameters used in Fig.~2 of the main text for various OAM and inductance values where $t=0$ corresponds to the moment where the electron is half-way through the loop. The solutions for the currents induced by an electron carrying one unit of OAM and circuits with inductance values of $0$, $0.05$, $0.1$, and $0.2$ pH are shown in Fig.~\ref{fig:inductionfig}(a). The same solutions are displayed in Fig.~\ref{fig:inductionfig}(b) over a zoomed in region of Fig.~{~\ref{fig:inductionfig}(a)} (highlighted in blue) corresponding to a longer time interval. We can clearly observe the influence of the inductor's presence in the circuit, i.e., the current varies less abruptly upon the electron's arrival and decays at a much slower rate after the electron exits the loop. More specifically, as the electron enters the loop, the inductor considerably reduces the circuit's current in such a way to minimize any variations in its magnetic flux. Unlike the case where there is no inductor, the current keeps on diminishing as the electron arrives midway through the loop in order to counter the fact that the induced EMF goes back to zero. The current starts to come back up once the electron starts leaving the loop and eventually starts decaying.

Fig~\ref{fig:inductionfig}(c) displays the induced current as the electron leaves the loop for OAM values of $\ell=1$, $5$, and $10$ and a circuit with an inductance of $0.1$ pH. As in the case where there is no inductor, the currents remain quantized and proportional to the electron's OAM value in addition to keeping observable values over a longer period of time.

\end{document}